
\input harvmac

\input epsf


\def\figin{\epsfcheck\figin}\def\figins{\epsfcheck\figins}
\def\epsfcheck{\ifx\epsfbox\UnDeFiNeD
\message{(NO epsf.tex, FIGURES WILL BE IGNORED)}
\gdef\figin##1{\vskip2in}\gdef\figins##1{\hskip.5in}
\else\message{(FIGURES WILL BE INCLUDED)}%
\gdef\figin##1{##1}\gdef\figins##1{##1}\fi}
\def\DefWarn#1{}
\def\figinsert{\goodbreak\midinsert}
\def\ifig#1#2#3{\DefWarn#1\xdef#1{fig.~\the\figno}
\writedef{#1\leftbracket fig.\noexpand~\the\figno}%
\figinsert\figin{\centerline{#3}}\medskip\centerline{\vbox{\baselineskip12pt
\advance\hsize by -1truein\noindent\footnotefont{\bf Fig.~\the\figno:} #2}}
\bigskip\endinsert\global\advance\figno by1}

\def\cpt{$\chi PT$}
\def\np{Nucl. Phys. }
\def\pr{Phys. Rev. }
\def\pl{Phys. Lett. }
\def\prl{Phys. Rev. Lett. }
\def \CO{{\cal O}}
\def \CM{{\cal M}}
\def\Re{{\rm Re~}}
\def\Im{{\rm Im~}}
\Title{\vbox{\baselineskip12pt\hbox{hep-ph/9403203, WIS-94/14/Feb-PH,
RU-94-24}}}
{{\vbox {\centerline{Missing (up) Mass, Accidental Anomalous
Symmetries,}
\centerline{and the Strong CP Problem}
}}}
\centerline{\it Tom Banks\foot{\tt banks@physics.rutgers.edu}}
\smallskip
\centerline{Department of Physics and Astronomy}
\centerline{Rutgers University, Piscataway, NJ 08855-0849, USA}
\vskip .1in
\centerline{\it Yosef Nir\foot{\tt ftnir@weizmann.bitnet}}
\smallskip
\centerline{Physics Department}
\centerline{Weizmann Institute of Science, Rehovot, 76100, ISRAEL}
\vskip .1in
\centerline{\it and}
\vskip .1in
\centerline{\it Nathan Seiberg\foot{\tt seiberg@physics.rutgers.edu}}
\smallskip
\centerline{Department of Physics and Astronomy}
\centerline{Rutgers University, Piscataway, NJ 08855-0849, USA}
\vskip .2in

\noindent
We reconsider the massless up quark solution of the strong CP problem.
We show that an anomaly free horizontal symmetry can naturally lead to a
massless up quark and to a corresponding accidental anomalous symmetry.
Reviewing the controversy about the phenomenological viability of
$m_u=0$ we conclude that this possiblity is still open and can solve the
strong CP problem.

\bigskip
\noindent
To appear in the Proceedings of The Yukawa Couplings and the Origins of
Mass Workshop.

\Date{2/94}

\newsec{Introduction}
In the present state of experimental particle physics, our strongest
clues to the nature of physics beyond the standard model are the various
fine tuning problems that are revealed by the standard model fit to
experimental data.  Surely the most pressing of these are the fine
tuning of the cosmological constant, and the gauge hierarchy problem.
Next in importance come the various small numbers associated with the
fermion mass matrices, and among these, the strong CP problem is (at
least numerically) the most striking.  Several mechanisms, of varying
degrees of plausibility, have been invented to account for the small
value of the QCD vacuum angle that is required to explain the observed
bounds on the neutron electric dipole moment.  In particular, it is
often suggested that if only a massless up quark were compatible with
the spectrum of hadrons, it would also provide a neat and economical
solution of the strong CP problem.

Recently a class of models which employed discrete symmetries and some
modest dynamical assumptions were proposed
\ref\lns{M. Leurer, Y. Nir and N. Seiberg, \np B398 (1993) 319;
hep-ph/9310320, RU-93-43, WIS-93/93/Oct-PH, \np in press.}
to explain the gross features of the quark and lepton mass matrices.
The simplest of these models automatically incorporated a discrete
symmetry which guaranteed that the up quark mass was zero to all orders
in perturbation theory. The discrete symmetry enforced an accidental
anomalous $U(1)$ symmetry on all renormalizable terms in the Lagrangian.
This symmetry remains unbroken in perturbation theory despite the fact
that the discrete symmetry is spontaneously broken.

The real and imaginary parts of the determinant of the quark mass matrix
are {\it a priori} independent parameters in the standard model.  In the
standard model, setting either or both of them to zero is unnatural.
The argument that the theory is more symmetrical for $\Re m_u=\Im
m_u=0$, and that the fine tuning is therefore natural, is spurious.  The
axial $U(1)$ ``symmetry'' of the standard model with one massless quark
is anomalous, and is really no symmetry at all.  In the absence of
further constraints on very high energy physics we should expect all
relevant and marginally relevant operators that are forbidden only by
this symmetry to appear in the standard model Lagrangian with
coefficients of order one.

In some of the models of \lns, it is natural to set the real and
imaginary parts of the up quark mass matrix to zero at high energy.  The
accidental anomalous $U(1)$ symmetry of these models guarantees that
both the real and imaginary parts are generated only by nonperturbative
QCD processes.  In section II of the paper we review the details of the
mechanism which leads to the accidental anomalous $U(1)$ symmetry, and
argue that this is essentially the unique way to generate such an
accidental symmetry.

To evaluate the viability of these models, we were led to reexamine the
controversy which has surrounded the question of whether a massless up
quark is consistent with low-energy hadron phenomenology.  In section
III of this paper we review the literature on this subject.  Fitting low
energy data to a first order chiral Lagrangian, one can extract the
``low-energy quark masses,'' $\mu_i$.  These should be distinguished
{}from the quark mass parameters, $m_i$ of the QCD Lagrangian at high
scale (say 1 TeV)
\ref\georgi{H. Georgi and I.N. McArthur, Harvard preprint HUTP-81/A011
(1981).}
\ref\choi{K. Choi, C.W. Kim and W.K. Sze, \prl 61 (1988) 794.}.
The distinction between them is second order in the $m$'s.  In
particular, $\mu_u$ receives an additive contribution of order $m_d m_s
\over \Lambda _{\chi SB} $ where $\Lambda _{\chi SB}\sim 1\ GeV$ is
some characteristic QCD energy scale, perhaps the fundamental scale
$4\pi f_{\pi}$.  Even if $m_u=0$, the parameter $\mu_u$ can be nonzero.
Its value is $\mu_u = \beta {m_d m_s \over \Lambda _{\chi SB}}$ where the
dimensionless coefficient, $\beta$, is estimated to be a number of order
one.  Unfortunately, with present technology we are unable to
calculate it reliably starting {}from QCD.  The low-energy data is
compatible with $m_u=0$ provided $\beta \approx 2$ which is of order one.
Therefore, $m_u=0$ seems to be a viable possibility.  Kaplan and Manohar
\ref\kapman{D.B. Kaplan and A.V. Manohar, \prl 56 (1986) 2004.}
tried to avoid a calculation of the parameter $\beta$ by extending the
low-energy analysis to second order in the $m$'s.  This led them to find
an ambiguity in the parametrization of the second order Lagrangian which
prevented them {}from determining the value of $m_u$.  Several authors
\ref\leut{H. Leutwyler, \np B337 (1990) 108.}
\ref\dw{J.F. Donoghue and D. Wyler, \pr D45 (1992) 892.}
\ref\dhw{J.F. Donoghue, B.R. Holstein and D. Wyler, \prl 69 (1992) 3444.}
\ref\mupiszero{K. Choi, \np B383 (1992) 58; \pl B292 (1992) 159.}
\ref\luty{M. A. Luty and R. Sundrum, \pl B312 (1993) 205; R. F. Lebed
and M. A. Luty, LBL-34779, UCB-PTH-93/28, hep-ph/9401232.}
tried to resolve this ambiguity and to determine $m_u$ by adding more
physical input.  Our understanding is that this ambiguity cannot be
resolved solely on the basis of low energy data.  Therefore, we conclude
that the possibility that $m_u=0$ is still open.

In section IV we take up the question of whether a vanishing up quark
mass really solves the strong CP problem.
In models with an accidental anomalous $U(1)$ symmetry, both real and
imaginary parts of the up quark mass are generated by nonperturbative
QCD dynamics. The message of section III
was that the real part so generated is, within our ability to calculate,
compatible with low energy data.  Is the same true of the imaginary
part?  This is a particular case of a general question first addressed
in 1979 by Ellis and Gaillard
\ref\ellis{J. Ellis and M.K. Gaillard, \np B150 (1979) 141.},
and since studied by a number of authors
\ref\shabalin{E.P. Shabalin, Sov. J. \np 28 (1979) 75; 31 (1980) 864.}
\ref\othercpcal{M. Dugan, B. Grinstein and L. Hall, \np B255 (1985) 413;
H. Georgi and L. Randall, \np B276 (1986) 241.}:
if the QCD vacuum angle is set to zero at some large scale $\Lambda_0$,
what will its low-energy value be?  We argue that in the present context
there is a very general operator analysis of this question.  The
analysis shows that CP violation in the high-energy theory can induce a
low-energy $\theta_{QCD}$ only via the action of a CP violating
irrelevant operator ${\cal O}$.  In this case, the same operator will
make a direct perturbative contribution to the neutron electric dipole
moment of order ${\Lambda_{\chi SB} \over \mu_u}\theta_{QCD}$ where
$\mu_u$ is the low energy up quark mass described above and
$\Lambda_{\chi SB}$ again denotes a typical QCD scale.  In other words,
if both the real and the imaginary part of $m_u$ vanish at high-energy,
nonperturbative strong CP violation will be smaller than perturbative
contributions to the neutron electric dipole moment. We conclude that
the models of \lns\ provide a phenomenologically viable and relatively
economical solution to the strong CP problem, in a framework which may
resolve many of the other puzzles of the fermion mass matrix in the
standard model.

In an appendix we present a warmup exercise for the calculation of direct
contributions to the neutron electric dipole moment in models with an
accidental anomalous $U(1)$ symmetry.

\newsec{$m_u=0$,  Naturally}

CP violation in the QCD Lagrangian arises {}from the terms:
\eqn\scpa{{\cal L}_{CP}=\theta\ {g^2\over32\pi^2}\ G^{\mu\nu}\tilde
G_{\mu\nu} -i\bar q ({\rm Im}\ M_q)\gamma_5 q}
where $M_q$ is the quark mass matrix.  There is only one independent CP
violating parameter in \scpa,
\eqn\scpb{\bar\theta=\theta+\arg\det M.}
Strong interactions, through their $\bar\theta$-dependence, lead to
an electric dipole moment for the neutron.
An estimate of this contribution, using current algebra, gives
\ref\cdvw{V. Baluni, \pr D19 (1979) 2227; R.J. Crewther, P. Di Vecchia,
G. Veneziano and E. Witten, \pl 88B (1979) 123.}
\eqn\scpc{{\cal D}_n=+3.6\times10^{-16}\ \bar\theta\ e\ {\rm cm}.}
The experimental bound on ${\cal D}_n$
\ref\dnexp{K.F. Smith {\it et al.}, \pl B234 (1990) 191;
I.S. Altarev {\it et al.}, JETP Lett. 44 (1986) 460.},
\eqn\scpd{{\cal D}_n\ <\ 1.2\times10^{-25}\ e\ {\rm cm}}
implies
\eqn\scpe{\bar\theta\ <\ 10^{-9}.}

In pure QCD the fine tuning required to satisfy \scpe\ is natural.  A
new symmetry, CP, is obtained for $\bar\theta=0$.  However, within the
Standard Model, this fine tuning is not natural, since the standard
model lagrangian with nonzero CKM phase $\delta$ is not CP invariant
even for $\bar\theta=0$. The experimental observation of CP violation in
the neutral $K$ system requires that $\delta$ be nonzero.

The above parametrization gives the erroneous impression that one fine
tuning, $m_u=0$, sets the up quark mass to zero and solves the strong CP
problem simultaneously. In fact, if we choose to absorb the vacuum angle
into the quark mass matrix, we can instead parametrize QCD by the real
positive values of $N_f - 1$ quark masses, and the real and imaginary
parts of the up quark mass\foot{In order for the eigenvalues of $\Re M$
to have the conventional meaning of quark masses, it is necessary to
have $\Im M$ proportional to the unit matrix. The difference between Re
$M$ and the quark masses in our convention is ${\cal O}(\bar\theta^2)$
and therefore negligible.}.  The bound on $\bar\theta$ can be
reexpressed as
\eqn\scpg{{\rm Im}\ m_u\ <\ 5\times10^{-12}\ GeV,}
which is about fourteen orders of magnitude below the electroweak
breaking scale.  Given this bound, the imaginary part of the up quark
mass makes a completely negligible contribution to hadron masses, and
chiral lagrangian fits to pseudoGoldstone boson masses are in fact
constraints only on the
independent, CP conserving, real part of the up quark mass.  The latter
is only constrained to be about five orders of magnitude below the
electroweak scale, and we will see that the data are consistent with it
being equal to zero.

In order to render the vanishing of both the real and the imaginary
parts of $m_u$ natural (with all other quarks massive), a continuous
$U(1)$ symmetry
\eqn\acci{\bar u \rightarrow e^{i\alpha} \bar u}
must be imposed.  It guarantees that the left handed field $\bar u$
couples only to the gauge fields.  Like the Peccei Quinn symmetries of
axion models, this $U(1)$ symmetry is anomalous and hence not an exact
symmetry of the full theory.  Therefore, it is unnatural to impose it on
a Lagrangian at a given energy scale, unless we can provide a reason why
symmetry violating physics at much larger energy scales does not violate
all conclusions based on the anomalous $U(1)$.  We conclude that, the
value of neither the real nor the imaginary part of the up quark mass is
natural in the standard model, although the imaginary part is certainly
the worst offender.

We should remark that in SUSY theories, the non-renormalization theorems
allow us to set the coupling of the up quark to the Higgs to zero.  In
such theories a massless up quark might appear to be technically
natural.  (Clearly, the symmetry \acci\ acts then also on the $\bar u$
squark.)  However, unless we arbitrarily impose the $U(1)$ symmetry on
the soft SUSY breaking terms (in particular a coupling of the $\bar u$
squark to a squark and a Higgs breaks the symmetry) radiative
corrections would lead to couplings of $\bar u$ to the Higgs and to an
up quark mass.  Since we do not expect the full theory to respect the
anomalous $U(1)$, the non-renormalization theorems do not help.

The only natural way to ensure an anomalous $U(1)$ symmetry is to make
it an accidental symmetry.  That is, it follows {}from another symmetry
plus renormalizability.  All terms which violate it are then irrelevant
in the renormalization group sense, and will have negligible effect if
the physics responsible for them is pushed off to a sufficiently high
energy scale.  Thus our strategy is as follows: We impose an anomaly
free symmetry $H$, which can be continuous or discrete, and require that
the most general renormalizable $H$ invariant Lagrangian exhibits the
accidental $U(1)$ symmetry \acci.

It is clear that in order to achieve this state of affairs, the symmetry
$H$ must act differently on $\bar u$ and on $\bar c$ and $\bar t$.  (It
can also act on the other light fields.)  It therefore satisfies the
definition of a {\it horizontal symmetry}.  As such we can immediately
apply some of the general analysis of horizontal symmetries presented in
\lns\ .

If $H$ acts only on $\bar u$, it is anomalous.  To cancel this anomaly
other fields in the theory must be in a complex representation of $H$.
Since they are massive, $H$ must be spontaneously broken.

We now examine how $H$ can be spontaneously broken.  The expectation
value of the single Higgs in the standard model cannot break $H$ \lns.
More precisely, a subgroup of $U(1)_Y \times H$ isomorphic to $H$ is
always unbroken ($U(1)_Y$ is hypercharge).  Similarly, in a two Higgs
version of the standard model with natural flavor conservation (such as
the supersymmetric standard model) $H$ cannot be broken by the
expectation values of the two Higgs fields $\phi_u$ and $\phi_d$ \lns.
To show that, note that in the absence of a $\phi_u \phi_d$ term in the
superpotential and in the soft breaking terms, the renormalizable theory
is invariant under the anomaly free $U(1)$ symmetry
\eqn\newsym{\eqalign {\bar u &\rightarrow e^{-3i\alpha} \bar u \cr
\bar d &\rightarrow e^{i\alpha} \bar d \cr
\bar s &\rightarrow e^{i\alpha} \bar s \cr
\bar b &\rightarrow e^{i\alpha} \bar b \cr
\phi_d &\rightarrow e^{-i\alpha} \phi_ d \cr}}
(all other fields are invariant) which guarantees the accidental $U(1)$
\acci.  However, its breaking by $\langle \phi_d \rangle$ leads to an
unacceptable Goldstone boson.  With the $\phi_u \phi_d$ term in the
Lagrangian we can again use $U(1)_Y$ to make both Higgs fields
invariant. Therefore, we must add more fields to the theory.

The simplest way to organize the analysis is to integrate out the extra
fields, which we assume to be heavier than the weak scale, and to
consider non-renormalizable terms added to the standard model
Lagrangian.  Typical non-renormalizable terms which could avoid the
previous no-go theorem are
\eqn\typnon{Q \phi_d \bar d \left({\phi_u \phi_d \over M^2}\right)^n
+Q \phi_u \bar u \left({\phi_u \phi_d \over M^2}\right)^n}
where $M$ is the scale of the fields which have been integrated out.
Such terms explicitly break the anomaly free continuous symmetry
\newsym\ to a discrete subgroup in the absence of a $\phi_u \phi_d$ term
in the Lagrangian.  We can also add a field, $S$, which is invariant
under the standard model gauge group, and whose vacuum expectation value
breaks $H$.  Then, terms like
\eqn\typnons{Q \phi_d \bar d \left({S \over M}\right)^n + Q \phi_u \bar
u \left({S \over M}\right)^n}
can lead to the entries in the mass matrix.  Rather low values of
$\langle S\rangle$ and $M$ are consistent with the constraints on flavor
changing processes \lns.  Terms of the form \typnon\typnons\ can be
generated by integrating out massive fermions of masses of order $M$
\ref\frog{C.D. Froggatt and H.B. Nielsen, \np B147 (1979) 277.}
and lead to interesting mass matrices for the quarks.

We conclude that the framework of \lns\ provides the unique natural way of
enforcing the accidental $U(1)$ symmetry, and hence $m_u=0$, to all orders
in perturbation theory. We must now examine the question of
whether a theory with such an accidental anomalous symmetry is
phenomenologically viable.

\newsec{Microscopic and Macroscopic Mass Matrices}

In this section we will review the controversy that has revolved around
the ``experimentally determined'' value of the up quark mass.  Early
estimates
\ref\weinberg{S. Weinberg, in {\it A Festschrift for I.I. Rabi},
ed. L. Motz (NY Acad. Sci., 1977) p. 185.}
of the pion and kaon masses in terms of quark masses in lowest order
\cpt\ led to the conclusion that the up quark mass could not be much
smaller than half the down quark mass.

In quantum field theory, the parameters in an effective Lagrangian are
scale dependent.  A microscopic theory more fundamental than the
standard model fixes them at some high-energy, e.g.\ of order $1\ TeV$.
In order to relate these to low-energy parameters we must understand how
the parameters renormalize in QCD.

To lowest order in chiral perturbation theory, the renormalization of
quark masses is multiplicative and flavor independent.  This is what
enabled Weinberg \weinberg\ to obtain renormalization invariant
predictions for high-energy quark mass ratios in terms of low-energy
meson masses.  Once we go beyond lowest order \cpt\ however, this is no
longer true.  In particular, the up quark mass receives an additive
renormalization proportional to $m_d ^* m_s^* $ via the mechanism of
\georgi\choi.  Their result may be phrased as the statement that a
renormalization group transformation {}from a scale $\lambda$ to
$\lambda-\Delta\lambda$ (for $\lambda\gg \Lambda_{\chi SB} \sim 1\ GeV$
where the theory is weakly coupled) leads to
\eqn\rengroup{\eqalign{m_i(\lambda-\Delta\lambda)= &m_i(\lambda)(1+\CO
(g^2(\lambda), m_j^2) ) \cr &+ {\det M_q^\dagger \over m_i^* }
\int_{\lambda^{-1}}^{(\lambda -\Delta\lambda)^{-1}} d\rho {1 \over
\rho^{N_f-3}} e^{-{8\pi^2\over g^2(\rho)}}g^n(\rho) \beta (1 +\CO
(g^2(\rho), m_j^2)) \cr} }
where the mass matrix $M_q$ has complex eigenvalues $m_i$ (the vacuum
angle $\theta$ has been rotated into $M_q$).  The second term arises
{}from small instantons and the dimensionless constant $\beta$ depends
on the number of colors $N$ and flavors $N_f$.

This term also depends on the regularization scheme through our choice
of integrating over the instanton scale size $\rho$ {}from $\lambda^{-1}$
to $(\lambda -\Delta\lambda)^{-1} $.  This ambiguity is very small.  It
is of order $\exp (-{8\pi^2\over g^2(\lambda)}) \ll 1 $.  Therefore,
even though $m_i(\lambda)$ is ambiguous, the limit $\lim_{\lambda
\rightarrow \infty} m_i (\lambda)$ is well defined and $M_q$ at high
energies is not ambiguous.

In renormalizing between two high-energy scales (e.g.\ 1 TeV and 100
GeV) the additive renormalization is accurately calculated in the dilute
instanton gas approximation, and is very small because it is
nonperturbative in the small high-energy value of the QCD coupling.
When the QCD coupling becomes strong this is no longer the case and the
best we can do is to estimate the additive contribution to the low
energy quark mass as ${m_d m_s \over \Lambda_{\chi SB}}$.  Therefore, if
$m_u=0$, $\mu_u \sim {m_d m_s \over \Lambda_{\chi SB}}$.  This estimate
is compatible with the low-energy quark mass ratios extracted {}from
lowest order chiral perturbation theory.  A more detailed comparison
requires a nonperturbative solution of QCD, and might in principle
(though probably not in practice since the effect pointed out in
\georgi\ and \choi\ is invisible in the quenched approximation) be
extracted {}from a lattice calculation.

To conclude this section, we emphasize that the high energy values of
the quark masses are, for all practical purposes, unambiguously
determined by potentially observable QCD Green's functions.  For
example, in the QCD sum rule approach to hadron phenomenology, it is
precisely these high energy values which are related to low energy
hadron parameters\foot{This implies that if our assertions about the
inequality of low energy and high energy quark mass ratios are correct,
then values for the quark masses which are extracted by combining sum
rules with chiral lagrangian results for mass ratios are incorrect in
principle.}.  Unfortunately, it is difficult to estimate the
uncontrollable systematic errors in sum rule calculations.  As far as we
have been able to determine, all such calculations are compatible with a
vanishing value for the high energy up quark mass.

\subsec{First order analysis}

To make these considerations more precise, let us outline the procedure
for actually extracting the high-energy quark masses {}from low-energy
data. The first step is to fit low-energy data to the low-energy chiral
Lagrangian (we use the notations of reference
\ref\leutwyler{H. Leutwyler, in {\it Perspectives in the Standard
Model}, eds. R.K. Ellis, C.T. Hill and J.D. Lykken (World Scientific,
1992), p. 97.}
\eqn\chirlag{\CL = {F_\pi^2 \over 4} \Tr\{ \partial_\mu U \partial^\mu
U^\dagger  +2B(\CM U^\dagger+\CM^\dagger U)\}.}
Using $SU(3)_L\times SU(3)_R$, $\CM$, which is in the $(3, \bar 3)$
representation, can be brought to the form
\eqn\chimat{\CM=\pmatrix{\mu_u e^{i\theta} & & \cr
                                            &\mu_d&\cr
                                            &&\mu_s\cr}}
where all the parameters are real.  The mass parameters $\mu_i$ can be
interpreted as the low-energy values of the high-energy quark masses
$m_i$.  The result of the fit is
\eqn\fitfirst{\theta < 10^{-9},\quad {\mu_u \over \mu_d} \sim 0.57, \quad
{\mu_d \over \mu_s} \sim 0.05.}

Now let us expand the $\mu$'s in the short distance $m$'s.  Using the
$SU(3)_L\times SU(3)_R$ transformation laws of the $m$'s we find that
for three flavors
\eqn\quarkex{\eqalign{
\mu_u& = \beta_1 m_u +\beta_2
{m_d^* m_s^* \over \Lambda_{\chi SB} } + \CO(m^3) \cr
\mu_d& = \beta_1 m_d +\beta_2
{m_u^* m_s^* \over \Lambda_{\chi SB} } + \CO(m^3) \cr
\mu_s& = \beta_1 m_s +\beta_2
{m_u^* m_d^* \over \Lambda_{\chi SB} } + \CO(m^3) \cr}} where $\beta_1$
and $\beta_2$ are dimensionless coefficients.  The $\mu$'s in \quarkex\
are complex and can be brought to the form \chimat\ using an
$SU(3)_L\times SU(3)_R$ transformation.  The additive renormalization
proportional to $\beta_2$ is the strong coupling version of the
instanton term in \rengroup\ and it represents the breaking of the axial
$U(1)$ by the anomaly.  In terms of the matrix $\CM$ and the underlying
quark mass matrix $M_q$, equation \quarkex\ can be written as
\eqn\quarkexchi{\CM=\beta_1 M_q +\beta_2{1\over \Lambda_{\chi SB} }
\det M_q^\dagger {1\over  M_q M_q^\dagger } M_q +\CO(M_q^3). }

The dimensionless coefficients $\beta_1$ and $\beta_2$ cannot be found
without a strong coupling calculation in QCD.  On general grounds we
expect them to be of order one.  They both suffer {}from ambiguities
resulting {}from the regularization scheme.  The only sense in which any
of them is small is in the large $N$ approximation, where $\beta_2$ is of
order $1 \over N$.

Assuming ${m_u\over m_d} \ll 1$, \quarkex\ leads for real $m$'s to
\eqn\quarkratio{{m_u \over m_d} = {\mu_u\over \mu_d} -
{\beta_2\over\beta_1^2} {\mu_s \over \Lambda_{\chi SB} }
+ \CO\left( ({\mu_i \over \Lambda_{\chi SB}} )^2\right).}
Therefore, $m_u \over m_d$ cannot be determined without knowledge of
$\beta_2\over\beta_1^2$.  If the dimensionless ratio $\beta_2\over
\beta_1^2$ is near 2, the up quark can be massless.  Even if this ratio
is not near 2, the ambiguity in $m_u \over m_d$ is significant.

\subsec{Second order analysis}

The previous discussion can be criticized on the basis that it is a
first order analysis in the $\mu$'s but it includes some second order
contribution in the $m$'s.  To make the analysis consistent one should
include all second order terms in the $m$'s and hence, go to second
order in the $\mu$'s.  The most general potential of second order in
$M_q$ can be written, using $\chi\equiv 2B M_q$, as
\eqn\secondpot{\eqalign{ V(U)=&\left\{{F_\pi^2\over4}\Tr\chi U^\dagger
 + r_1 [ \Tr \chi^\dagger U\chi^\dagger U - (\Tr \chi^\dagger U)^2 ]
\right\} + {\rm h.c.} \cr
 +& r_2 [ \Tr \chi^\dagger U\chi^\dagger U + (\Tr \chi^\dagger U)^2 ]
 + {\rm h.c.} \cr
+& r_3 (\Tr  \chi^\dagger U)(\Tr  \chi U^\dagger) +\CO(\CM^3) .
\cr} }
(The dimensionless coefficients $r_1$, $r_2$ and $r_3$ are linear
combinations of $L_6$, $L_7$ and $L_8$ of \leutwyler.)  Remembering that
$M_q$ is in the $(3,\bar 3)$ representation of the $SU(3)_L\times
SU(3)_R$ flavor symmetry, we learn that the terms multiplying $r_1$,
$r_2$ and $r_3$ are in the representations $(3,\bar 3)$, $(\bar 6,6)$
and $(8,8)$, respectively.  Since the two terms in the curly brackets in
\secondpot\ transform as one irreducible representation, there must be a
redundancy in the parametrization.  Indeed, using the identity of $3
\times 3$ matrices
\eqn\kmidentity{\det A - {1\over 2} A [(\Tr A)^2 - \Tr A^2] +A^2 \Tr A
-A^3=0}
applied to $A=\chi^\dagger U$ we find
\eqn\anoiden{\det\chi^\dagger \Tr {1\over \chi^\dagger} U^\dagger  +
{1\over 2} \Tr \chi^\dagger U\chi^\dagger U- {1\over 2} (\Tr
\chi^\dagger U)^2  =0.}
Therefore, the shifts
\eqn\kmshift{\eqalign{\chi \rightarrow & \chi + { 8a\over F_\pi^2} (\det
\chi^\dagger) {1\over \chi^\dagger} \cr
r_1 \rightarrow &r_1 + a \cr }}
with an arbitrary complex number $a$ change the potential \secondpot\ by
terms of order $M_q^3$, which are neglected.

Because of this redundancy in the parametrization of the effective
Lagrangian, the mass matrix $M_q$ cannot be determined by fitting the
experimental data to \secondpot.  At best, we can determine $M_q$ up to
an arbitrary shift by ${ aB\over F_\pi^2} (\det M_q^\dagger) {1\over
M_q^\dagger}  $.

The ambiguity \kmshift\ is reminiscent of the expression for $\CM$ in
terms of $M_q$ \quarkexchi.  The functional similarity stems {}from the
fact that the axial $U(1)$ is broken preserving only the $SU(3)_L\times
SU(3)_R$ symmetry.  However, these two effects are different.  Equation
\quarkexchi\ relates the short distance mass $M_q$ to the long distance
mass $\CM$ defined by a fit to a first order Lagrangian.  The non-linear
term in this equation expresses a renormalization effect.  The ambiguity
\kmshift\ expresses a redundancy of the second order low-energy
description which prevents us {}from using only the low-energy chiral
Lagrangian to determine $M_q$.

The distinction between these two effects is more clear in the
``trivial'' two flavor case.  The identity analogous to \anoiden\ in
this case is
\eqn\twoflai{\Tr \psi^\dagger U= \det \psi^\dagger \Tr {1 \over
\psi^\dagger} U^\dagger}
where $\psi$ is an arbitrary $2\times 2$ matrix.  It shows that the
matrix $\chi$ in the potential $\Tr \chi^\dagger U + c.c$ is ambiguous.
First, it is clear that only the eigenvalues of $\chi$ are physical.
Second, using equation \twoflai\ the transformation
\eqn\ambichit{\chi \rightarrow \chi + \psi - \det \psi^\dagger {1 \over
\psi^\dagger} }
leaves the potential invariant and allows us to set one of the
eigenvalues of $\chi$ to zero\foot{Note that therefore, in the two
flavor case there cannot be any CP violation in the first order chiral
Lagrangian.}.  The non-perturbative contribution to $\CM$ arises at
first order in $M_q$.  It is the second term in
\eqn\quarkexchit{\CM=\beta_1 M_q +\beta_2 \det M_q^\dagger {1\over
M_q M_q^\dagger } M_q +\CO(M_q^2). }
Clearly, the redundancy in the parametrization has nothing to do with
the change in the strength of this effect (changing $\beta_2$).  The
only common thing about the ambiguity \ambichit\ and the additive
renormalization \quarkexchit\ is that they both allow for $m_u=0$ to be
compatible with low energy data.

Some authors \leut\dw\dhw\ have suggested the use of more input about
low energy hadron physics to resolve this ambiguity and determine $M_q$.
We would like to make clear what bothers us about these attempts.  These
authors fix the ambiguity by insisting that the coefficients $\chi$ in
$Tr \chi U^\dagger$ in the chiral Lagrangian are proportional to the
bare quark masses, i.e.\ by setting what we have called $\mu_u \propto
m_u$.  They then attempt to compute the other coefficients in the
Lagrangian (implicitly) with this choice of Kaplan-Manohar ``gauge.''
However, their computations make no reference to this particular choice.
We could equally well consider them to be determinations of the
couplings in some other ``gauge.''  And since their computations involve
relatively low energy hadronic physics there is no reason to think that
they have anything to do with the special parametrization in which the
low energy and high energy up quark masses are proportional to each
other.  Other approaches of resolving the ambiguity use the large $N$
expansion.  While this may be a valid approach, one must beware of
considerations that are justified only in this limit.  The whole issue
is of higher order in $1\over N$ and therefore cannot be settled at the
leading order in this expansion.

Finally, we note that the argument is sometimes made that the
consistency of the quark masses extracted from first order chiral
perturbation theory analysis of baryons and mesons, suggests that the
scale dependent additive renormalization which we have discussed, is
small.  That is, the additive renormalization should give an effective
upquark mass proportional to $m_d m_s$ in both the baryon and meson
effective lagrangians, but the coefficients would not be the same.
We have examined the extraction of quark masses from baryon masses
\ref\gl{J. Gasser and H. Leutwyler, Phys. Rep. 87 (1982) 77.}
\dw\ and believe that it suffers from large uncertainties.
The results are sensitive to nonanalytic
corrections $(\propto m_q^{3/2})$, and to model dependent calculations of
electromagnetic contributions.  They are not precise enough to rule out
the possibility that $m_u = 0$.

{\it We conclude that at the level of precision (order of magnitude) of
nonperturbative QCD calculations available to us at present, low-energy
phenomenology is completely compatible with a vanishing value of the
high-energy up quark mass}.

Only a nonperturbative calculation in QCD can prove or disprove the
phenomenological viability of $m_u=0$.  Therefore, in view of the recent
progress in numerical methods in lattice gauge theory, we would like to
encourage a detailed analysis of the possibility of a massless up quark
by these methods.  We emphasize however that the additive
renormalization \rengroup, vanishes in the quenched approximation.

\newsec{Does $m_u=0$ Solve the Strong CP Problem?}

As we have seen in section II, models of the type studied by \lns\ can
incorporate discrete symmetries which guarantee that both the real and
imaginary parts of the up quark mass are zero (at $1\ TeV$) while CP
violation in the Cabibbo-Kobayashi-Maskawa matrix is allowed.  The
discrete symmetries enforce an accidental anomalous $U(1)$ symmetry on
the Lagrangian.  If all physics not explicitly included in the
Lagrangian comes {}from energy scales substantially higher than $1\ TeV$
then this accidental symmetry guarantees that the only nonvanishing
contributions to the up quark mass come {}from nonperturbative QCD
effects.

We have argued above that in our present state of impotence with regard
to precise nonperturbative calculations in QCD, the real part of the up
quark mass generated in this manner appears to be consistent with hadron
phenomenology.  The purpose of the following discussion is to determine
whether an analogous conclusion can be made for the imaginary part of
the up quark mass.

This task is made easier by the observation that violations of the
anomalous $U(1)$ symmetry are only significant at scales below $1\ GeV$.
Thus if we construct an effective field theory for scales just above $1\
GeV$, it must obey the $U(1)$ symmetry.  As a consequence, there will be
no CP violating terms in the renormalizable part of this effective
Lagrangian.  The imaginary part of the low-energy up quark mass will
thus be generated by a combination of nonperturbative QCD effects and CP
violating irrelevant operators in the effective field theory.  Let
${\cal O}_d$ be such an operator, of dimension $d$.  It will give a
contribution to the imaginary part of the up quark mass of order $C {m_d
m_s \over \Lambda_{\chi SB}}({\Lambda_{\chi SB}\over M}) ^{d -4}$, where
${C\over M^{d-4}}$ is the coefficient of ${\cal O}_d$ in the effective
Lagrangian, $M$ is the heavy scale at which CP violating effects
originate (e.g. the mass of the $W$ boson), and $\Lambda_{\chi SB}$ is,
as always, a typical QCD scale of order $1\ GeV$.  The consequent
contribution to the neutron electric dipole moment is of order $C {m_d
m_s \over \Lambda_{\chi SB}^3} ({\Lambda_{\chi SB}\over M})^{d -4}$.

The operator ${\cal O}_d$ also makes a direct contribution to the
neutron electric dipole moment of order $C {1\over \Lambda_{\chi SB}} (
{\Lambda_{\chi SB} \over M})^{d -4}$, which is larger than that coming
{}from the ``induced strong CP violation'' by a factor of order $
{\Lambda_{\chi SB}^2 \over m_d m_s }$.  {\it Thus, if the direct
contributions to the neutron electric dipole moment coming {}from
perturbatively induced CP violating irrelevant operators are within
experimental bounds, there will be no strong CP problem}.  Perturbative
CP violating effects might put significant constraints on particular
models with new physics at the $TeV$ scale. We will not study these
direct contributions in detail here (except for a short discussion of
the standard model in the appendix).  What we have shown is that if these
constraints are satisfied, we need not worry about strong CP violation.

\centerline{\bf Acknowledgements}

It is a pleasure to thank A. Cohen, A. Dabholkar, M. Dine, K.
Intriligator, D. Kaplan, A. Nelson, J.  Polchinski, S.  Shenker and E.
Witten for useful discussions.  Y.N.  wishes to thank the Rutgers group
for its hospitality.  This work was supported in part by DOE grant
DE-FG05-90ER40559. YN is an incumbent of the Ruth E.  Recu Career
Development chair, and is supported in part by the Israel Commission for
Basic Research, by the United States -- Israel Binational Science
Foundation (BSF), and by the Minerva Foundation.

\appendix{A}{Direct Contributions to the Neutron Electric Dipole Moment}

Here we present an operator analysis of direct perturbative
contributions to the neutron electric dipole moment in the standard
model with an accidental $U(1)$ symmetry.  This is a warmup for a full
calculation in models of the sort studied in \lns .

The lowest dimension CP violating operators that might be relevant here
are the dimension five chromoelectric dipole moments of the light
quarks.  In models with an accidental $U(1)$ symmetry, no dipole moments
involving an up quark are allowed in the effective Lagrangian above the
QCD scale.  It is also worth reiterating
\ref\signets{A. De R\'ujula, M.B. Gavela, O. P\`ene and F.J. Vegas,
\np B357 (1991) 311.}
that all dipole moments are suppressed by factors of light quark masses
because they violate the nonabelian chiral symmetries of massless QCD.
As a consequence their contributions to the neutron electric dipole
moment are {\it a priori} smaller than chirally allowed dimension six
operators, including the three gluon operator
\ref\russian{S. Weinberg, \prl 63 (1989) 2333.},
and the {\it axial polyp} \signets\ operators:
\eqn\scpi{\eqalign{{\cal L}_u^3=&\Delta_3^u\ {\cal P}_{3-}^u,\cr
{\cal P}_{3-}^u=&i\bar u\gamma^\mu\{G_{\mu\nu},\overrightarrow D^\nu
\}\gamma_5 u,\cr}}
There is a similar term, ${\cal L}_u^1$ with a photon field $F_{\mu\nu}$
replacing the gluon field $G_{\mu\nu}$ in \scpi.  In the standard model,
the largest contributions to quark electric dipole moments at a given
scale actually come {}from ``fusing'' together one of these polyp
operators and a quark mass term via the exchange of a gluon right at the
infrared cutoff scale.  In many previous analyses of CP violation, the
electric dipole moment operators were considered to be more important
than, or on equal footing with, the polyp operators.  We believe that
such arguments rest on the dubious procedure of sticking a
``constituent'' quark mass into an effective lagrangian.  The purpose of
an effective lagrangian calculation is to make a clean separation
between physics at different scales, and in particular between weakly
coupled high energy physics, and strongly interacting QCD.  We do not
believe that it is consistent with the ``rules of the game'' to
calculate the coefficients of an effective lagrangian (without the use
of a computer) at scales at which QCD is strongly coupled.  Thus,
constituent quark masses should not appear as coefficients in an honest
effective lagrangian.  Rather, we should work with operators normalized
at a scale where QCD is still weakly coupled and estimate their hadronic
matrix elements by dimensional analysis or some more sophisticated
hadronic ``model.''  In the framework of a such a philosophy, the direct
contributions of CP violating polyps are more important than
chromoelectric dipole moments of light quarks.

At the order of magnitude level then, the calculation of the neutron
electric dipole moment in the models of \lns\ probably reduces to the
calculation of $\Delta_3^q$ for the various quarks.  As a warmup
exercise for this calculation we will present an estimate for the
$\Delta_3^u$ coefficient within the Standard Model.

As a single $W$-propagator cannot introduce CP violation, we are led to
consider the diagrams of order $g_s\alpha_w^2$ shown in Fig. 1 - 3.  The
dependence of such diagrams on the quark sector parameters is of the
form
\eqn\cala{\eqalign{\sum_{i,k=1}^3&\sum_{j=1}^3 f_1(m_i)f_2(m_j)f_3(m_k)
{\rm Im}[V_{ui}V^\dagger_{ij}V_{jk}V^\dagger_{ku}]\cr =
\sum_{i,k=1}^3&\sum_{j=1}^3f_1(m_i)f_2(m_j)f_3(m_k)\
J\sum_{m,n=1}^3\epsilon_{1jm}\epsilon_{ikn}.\cr}}
$J$ is the CP violating invariant measure of Jarlskog
\ref\jarlskog{C. Jarlskog, \prl 55 (1985) 1039.}.
As only left handed fields have charged current interactions, there can
be no odd number of helicity flips in the loops, so the $f_i$ functions
are functions of quark squared-masses, $m_q^2$.  Assume that we had
replaced $f_2(m_j)$ by a function independent of the quark masses.
Since $\sum_j\epsilon_{1jm} =0$, such a quantity would not contribute to
the coefficient of CP violating operators. Similarly, if either of
$f_1(m_i)$ and $f_3(m_k)$ were constant, there would be no contribution.
Therefore, we can rewrite the above quantity by adding and subtracting
contributions which do not contribute to CP violating terms.  The result
is:
\eqn\calb{J\{[f_1(m_s^2)-f_1(m_b^2)][f_2(m_c^2)-f_2(m_t^2)]
[f_3(m_d^2)-f_3(m_b^2)]-
(m_s^2\leftrightarrow m_d^2)\}.}

\ifig\fvone{First Feynman diagram for induced $\theta$}
{\epsfxsize3.0in\epsfbox{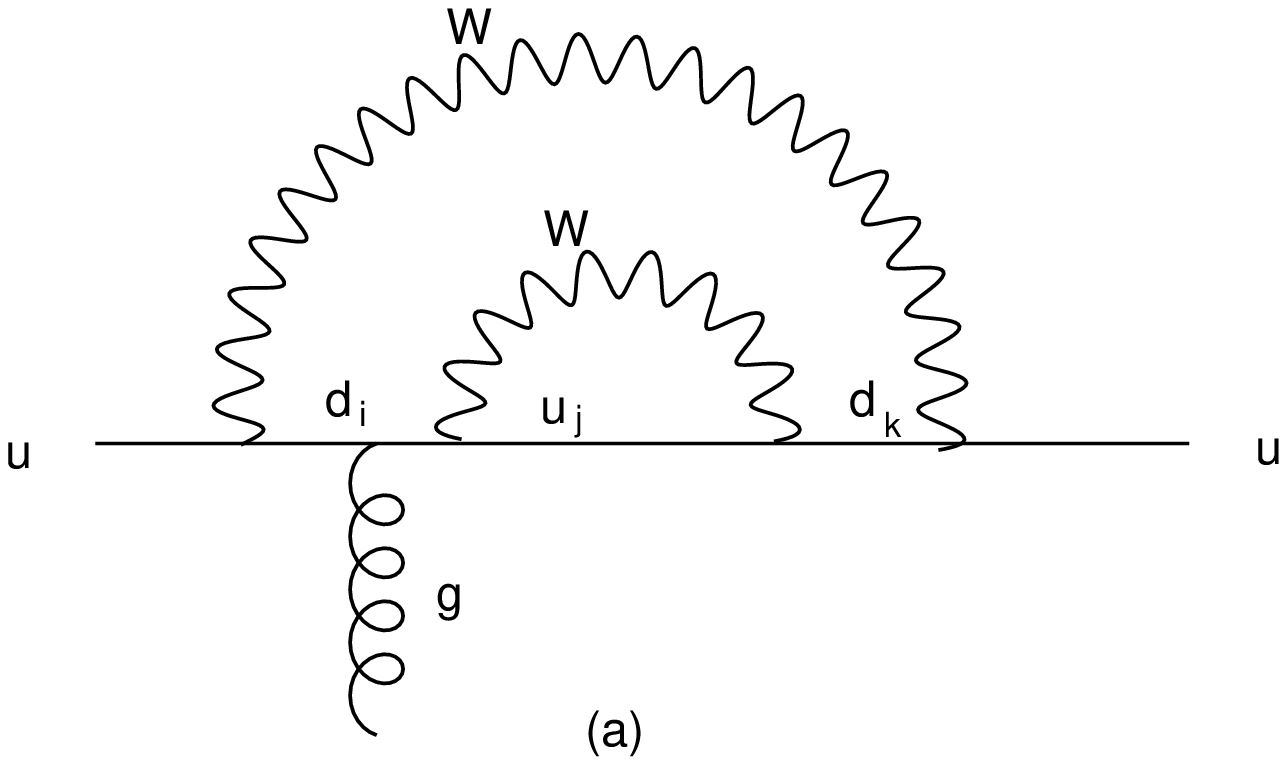}}

\ifig\fvtwo{Second Feynman diagram for induced $\theta$}
{\epsfxsize3.0in\epsfbox{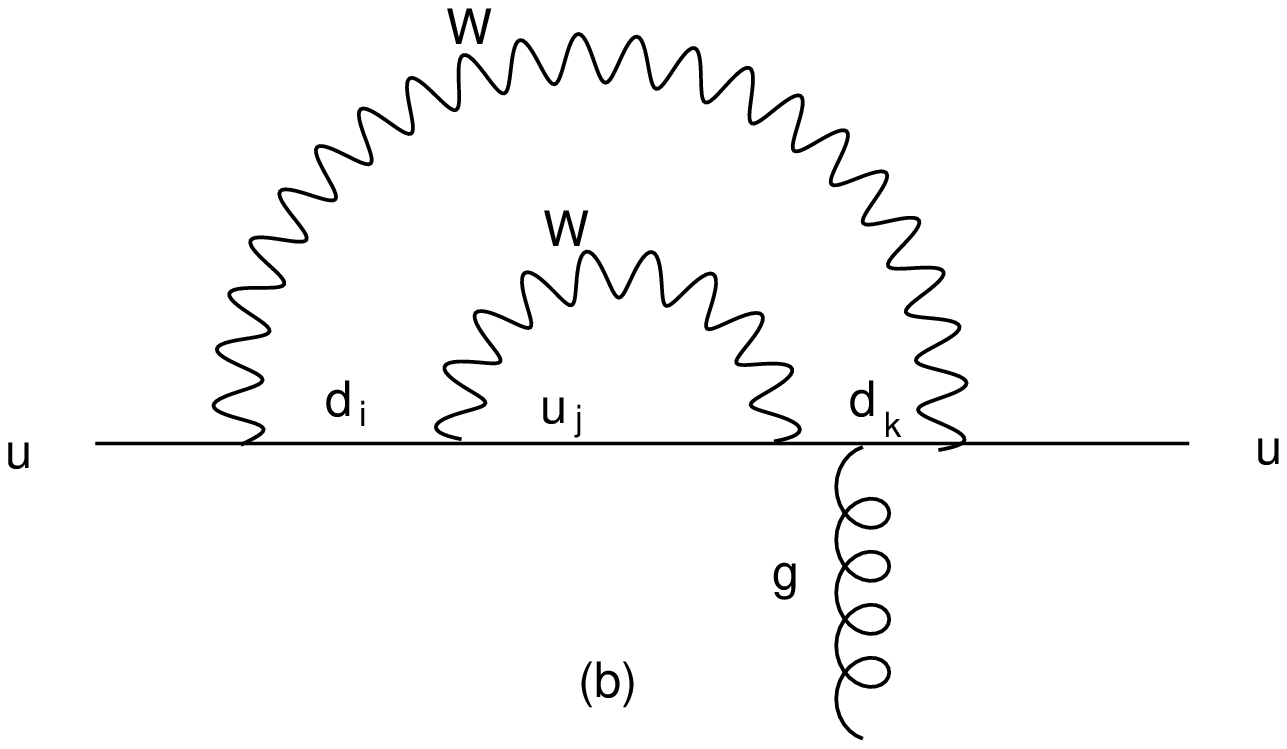}}

\ifig\fvthree{Third Feynman diagram for induced $\theta$}
{\epsfxsize3.0in\epsfbox{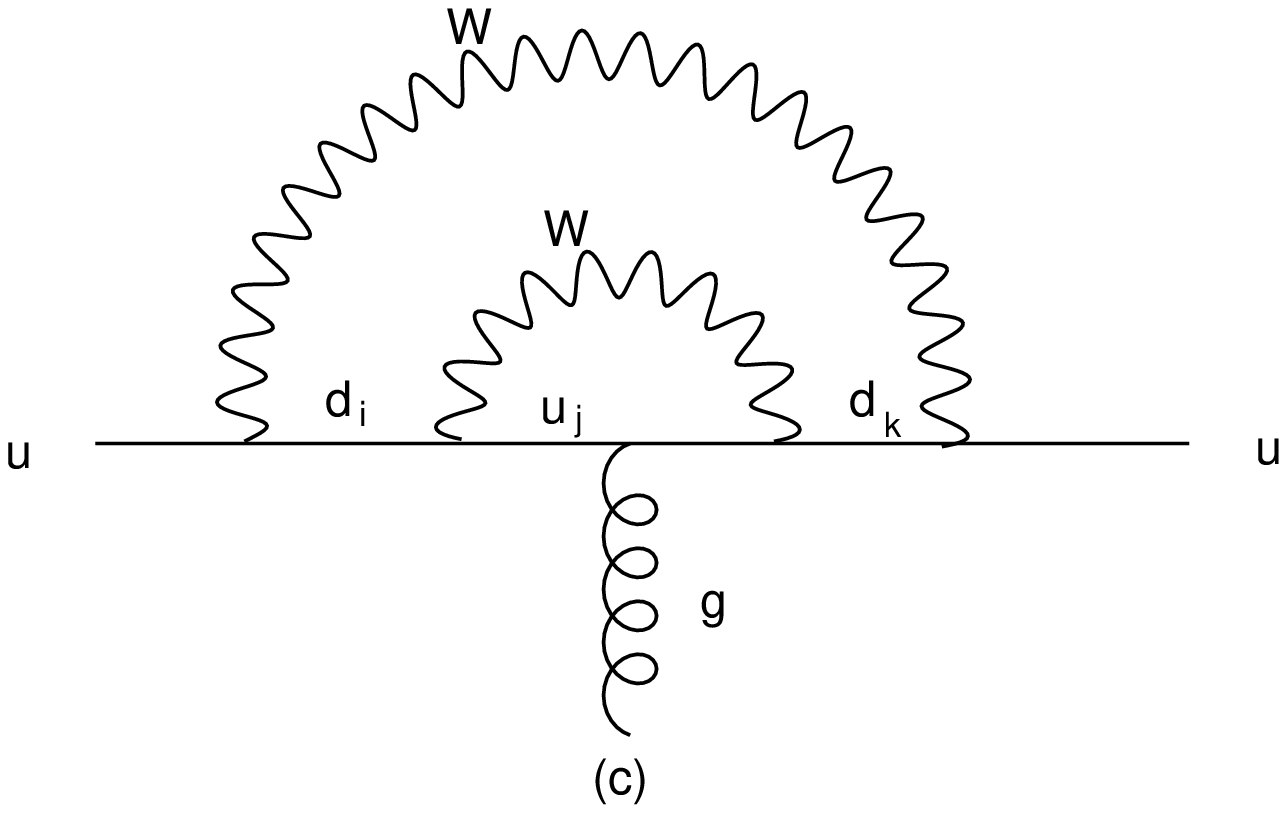}}

Next, let us concentrate on the inner loop of diagrams 1(a) and 1(b). We assume
that the momentum $p$ entering this loop is small on the scale of $m_W$.
(This assumption will be justified below.) Then such a loop contributes
\eqn\nonum{\eqalign{L_1\sim&\int d^4l{1\over (l+p)^2+m_W^2}
\left({\rlap/l\over l^2+m_c^2}-{\rlap/p\over l^2+m_t^2}\right)\cr
=&(m_t^2-m_c^2)\int d^4l {\rlap/l\over
((l+p)^2+m_W^2)(l^2+m_c^2)(l^2+m_t^2)}.\cr}$$ We next expand in powers
of $p$. The zeroth order term does not contribute because of Lorentz
invariance. The first order term is
$$L_1\sim(m_t^2-m_c^2)\rlap/p\int{d^4l\ l^2\over
(l^2+m_W^2)^2(l^2+m_c^2)(l^2+m_t^2)}.  }
We can scale the integral by $m_t^2$, assumed to be of order $m_W^2$.
The integral is finite, and consequently $L_1\sim\rlap/p$.  If $m_t$
had been smaller than $m_W$, there would have been an additional
suppression factor of order ${m_t^2} \over m_W^2 $.

Thus we can replace the inner loop with an insertion of a
$\rlap/p$-operator, where $p$ is the momentum entering this loop.  In
this approximation, the external loop of the diagram with the gluon on
the left (Fig. 1) gives
\eqn\anotherno{\eqalign{L_2&\sim\int
d^4l[(\rlap/l+\rlap/q)\gamma_\mu\rlap/l
\rlap/l\rlap/l]{1\over((l-k)^2+m_W^2)}
\left\{\left[{1\over(l^2+m_s^2)}-{1\over(l^2+m_b^2)}\right]\right. \cr
\times&\left.\left[{1\over((l+q)^2+m_d^2)(l^2+m_d^2)}-
{1\over((l+q)^2+m_b^2)(l^2+m_b^2)}\right]
-(m_s^2\leftrightarrow m_d^2)\right\},\cr}}
where $q$ is the incoming gluon momentum and $k$ is the incoming
$u$-quark momentum.  The diagram with the gluon on the right (Fig. 2)
gives a similar contribution in which the coefficient of
$(\rlap/l+\rlap/q)\gamma_\mu\rlap/l \over {(l - k)^2 + m_W^2}$ is
modified by simultaneously exchanging $m_d$ with $m_s$ and $l$ with
$l+q$.  If we now add these diagrams and perform the explicit
subtraction of terms with $m_d \leftrightarrow m_s$, we obtain:
\eqn\nona{\eqalign{L_2\sim&(m_s^2-m_d^2)(m_b^2-m_s^2)(m_b^2-m_d^2)
\int{d^4l\ [l^2 - (l+q)^2](\rlap/l+\rlap/q)
\gamma_{\mu}\rlap/l\over((l-k)^2+m_W^2)
((l+q)^2+m_b^2)(l^2+m_b^2)}\cr \times&{1\over
((l+q)^2+m_s^2)(l^2+m_s^2)((l+q)^2+m_d^2)(l^2+m_d^2) }.\cr}}
We are interested in the terms linear in both $k$ and $q$.  Thus we
expand to first order in $k$ and $q$. As $L_2$ is a highly convergent
integral even when we neglect the momentum dependence of the $W$
propagator, we were justified in assuming that the momentum entering the
internal loop was small.  We obtain a factor $m_W^{-4}$, multiplying a
finite integral.  Scaling the loop momentum by $m_b$ and expanding in
$({m_d \over m_s})^2$ and  $({m_s \over m_b})^2$ we get
\eqn\nonub{L_2\sim {m_s^2 \over m_W^4}
\int{d^4l\ l\cdot k l\cdot q \rlap/l\gamma_{\mu} \rlap/l \over
(l^2 + 1 )^2 (l^2 + ({m_d \over m_b})^2)^2 (l^2 + ({m_s\over m_b})^2
)^2}.}
This leads to\foot{For the diagram in Fig. 3,
the inner loop can be replaced with
a gluon vertex $g_s\gamma_\mu$. It contributes to $\Delta^u_3$
with the same order of magnitude as diagrams (a) and (b).}
\eqn\nonuc{\Delta^u_3\sim \left( {\alpha_w\over
\pi}\right)^2\ J\  {m_s^2\over m_W^4}.}

Using this result, it is easy to estimate the induced $\theta$.  We can
``fuse'' the polyp operator with the up quark mass (which in our case is
induced dynamically) to find
\eqn\indimup{\Im m_u \sim (\Re m_u) \Lambda_{\chi SB}^2
\Delta^u_3 \sim J ({\alpha_w\over \pi})^2 (\Re m_u) {\Lambda_{\chi SB}^2
m_s^2 \over m_W^4} } which agrees with the estimate of \ellis.  Note
that with the extra suppression factor coming {}from $L_1\sim {m_t^2
\over m_W^2}\rlap/p$ when $m_t < m_W$, our estimate \indimup\ would
agree with that of Shabalin \shabalin.

\listrefs
\end